# A TECHNICAL REVIEW ON COMPARISON AND ESTIMATION OF STEGANOGRAPHIC TOOLS

## Ms. Preeti P. Bhatt, Mr. Rakesh R. Savant

**Abstract—**Steganography is technique of hiding a data under cover media using different steganography tools. Image steganography is hiding of data (Text/Image/Audio/Video) under a cover as Image. This review paper presents classification of image steganography and the comparison of various Image steganography tools using different image formats. Analyzing numerous tools on the basis of Image features and extracting the best one. Some of the tools available in the market were selected based on the frequent use; these tools were tested using the same input on all of them. Specific text was embedded within all host images for each of the six Steganography tools selected. The results of the experiment reveal that all the six tools were relatively performing at the same level, though some software performs better than others through efficiency. And it was based on the image features like size, dimensions, and pixel value and histogram differentiation.

## 1. INTRODUCTION

In this current generation, technology is developing at fast step and new enlargements are made, safety is of utmost precedence.  The data has to be secure so that only authorized user can use the data. Lots of data are share through internet and transmitted on internet from one network to another. So the privacy of data is main concern of the sender. So the solution is to send the message in a private and secret way and only understand by receiver.

For securing the data two most important techniques are employed i.e. cryptography and steganography.

Previously only sender and receiver is can encrypt and decrypt the message by cryptography technique. Where cryptographic key is help to decrypt the message send by authorized sender. But it has some drawbacks that the person not participate in exchange of information and one can identify the message was hidden and can be decode by other person. To overwhelm this restriction the technique of steganography was presented [1].

The word "steganography" belongs to "Greek" language. In Greek it mean hidden writing or enclosed writing. The Steganography is technique to hide the data or message in any technical medium such as audio, video, image. It is better than cryptography as message is covered by image or audio or video. And have several advantages like: person does not know about hidden data. And it could be decrypt by authorize person from the send medium. Authorized person need to know the secret key and method to decode the data. No one can change the data except authorized user hence the safety and the consistency of data transmission also improved with invention of steganography [2].

Steganography methods have been used for long periods. Steganography has been broadly used in ancient times, especially before cryptographic systems were developed. The first known submission dates back to the earliest Greek times, Where message were painted on shaved heads and then let the hair to be grow, information remains unseen and secure. Second method used by them is wax table as cover source. On underlying wood the text was written and message was enclosed with the new wax layer. The blank tablets is passed and to be examined without any problem[3].

During Second World War, the invisible ink to write message on paper and transfer the message as a blank paper, so no one can read the data or identified the data. The special substance was used to decode the message. These liquids is heated and darken, so text can be visible by human eyes. Substance used such as urine, milk, vinegar and fruit juice.

"Ave Maria" cipher is another invention in Steganography. The book have lots series of tables and each table consist of list of words, one per letter. To encode the information, the message letter are replaced by respective words. If table are in order form then, one table per letter, then it should be like prayer.

All of these have common approach, the information are passed through or covered through physical object and then sent. The cover message is just an interruption, and could be anything of the numerous differences on this theme, only the pure information of the cover message is transmitted, so none of them will work for electronic transfer. However, there is plenty of room to hide private data in a not-so-private information. It just takes cleverness.

## 1.1.  Application of Steganography
Steganography is having various applications amongst them few are listed follow:

1.    Feature Tagging
2.    Copyright Protection
3.    Medical
4.    Secret Communication
5.    Use by terrorists
6.    Digital Watermarking

## 1.2.  Classification of Steganography
Secrets can be hidden inside all sorts of cover information. The following formula provides a very generic description of the pieces of the stenographic process: The private data is covered inside certain layer. The formula on which steganography works is as follows

**outer medium + secret data + private key=  stego medium**

Where outer medium is a file in which we will hide the private data, which may be encoded using private key. The final result is stego medium- which is same type as outer medium.[4]

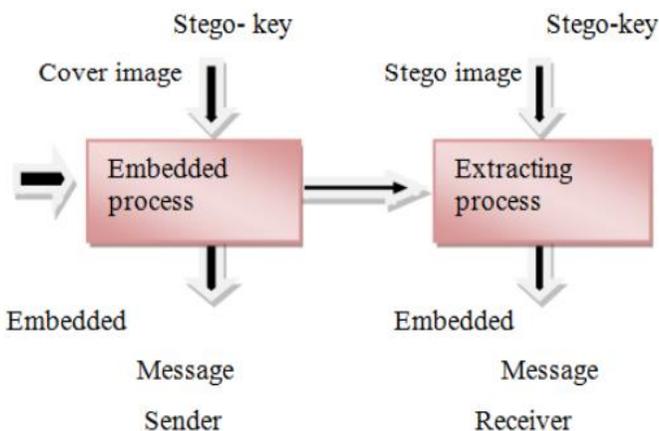

Figure 1: General Block Diagram of Steganography

There are four techniques to implement steganography:
1.    Using text.
2.    Using images.
3.    Using audio files.
4.    Using video files.

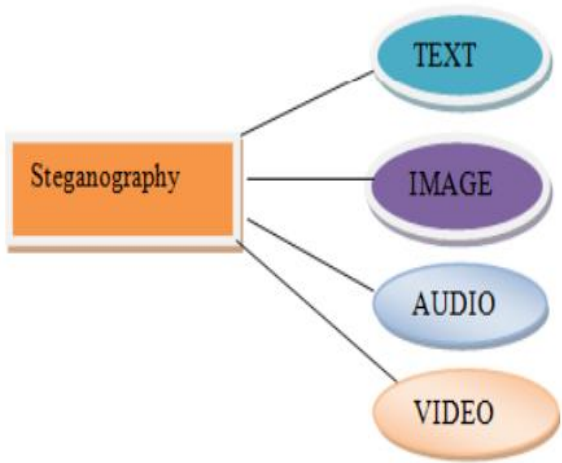

**Figure 2: Steganography Media**

### A.  Text Steganography:

Text steganography can be divided in to three basic categories - random, format-based, and statistical generation and linguistic method.

Format based technique format the physical text method to hide the information. In this approach it updates the existing text to hide the stenographic text. Some format based methods are used in text steganography are Insertion of spaces, deliberate misspellings distributed throughout the text, resizing. Bennett has proven that those method are only used to trick most of the human eyes but can't trick computer system. Random and statistical generation methods is producing cover text according to the statistical properties and it is based on character sequences and words sequences. The wrapping of information within character sequences is inserting the information to be appeared in random order of characters. This order must appear to be random to anyone who intercepts the message[5].

A second approach for character generation is to take the statistical properties of word-length and letter frequency inorder to create "words" (without lexical value) which will appear to have the same statistical properties as actual words in given language. The hiding of information within word sequences, the actual dictionary items can be used to encode one or more bits of information per word using a codebook of mappings between lexical items and bit sequences, or words themselves can encode the hidden information.

The final category is linguistic method which specifically considers the linguistic properties of generated and modified text, frequently uses linguistic structure as a place for hidden messages. In fact, steganographic data can be hidden within the syntactic structure itself.

### Example:

Sender sends a series of integer number (Key) to the recipient with a prior agreement that the secret message is hidden within the respective position of subsequent words of the cover text. For example the series is "1, 1, 2, 3, 4, 2, 4" and the cover text is "A team of five men joined today". So the hidden message is "Atfvoa". A "0" in the number series will indicate a blank space in the recovered message. The word in the received cover text will be skipped if thenumber of characters in that word is less than the respective number in the series (Key) which shall also be skipped during the process of message unhide.

### B. Image Steganography:

The most widely used technique today is hiding of secret messages into a digital image. This steganography technique exploits the weakness of the human visual system (HVS). HVS cannot detect the variation in luminance of color vectors at collection of color pixels. The individual pixels can be represented by their optical higher frequency side of the visual spectrum. A picture can be represented by a characteristics like brightness, Chroma etc. Each of these characteristics can be digitally expressed in terms of 1s and 0s[5].

### For example:

A 24-bit bitmap will have 8 bits, representing each of the three color values (red, green, and blue)at each pixel. If we consider just the blue there will be 2 different values of blue. The difference between 11111111 and11111110 in the value for blue intensity is likely to be undetectable by the human eye. Hence, if the terminal recipient of the data is nothing but human visual system (HVS) then the Least Significant Bit (LSB) can be used for something else other than color information.

## C.  Audio Steganography:

Steganography in general, relies on the imperfection of the human auditory and visual systems. Audio steganography takes advantage of the psychoacoustical masking phenomenon of the human auditory system [HAS]. Psychoacoustical or auditory masking property renders a weak tone imperceptible in the presence of a strong tone in its temporal or spectral neighbourhood. This property arises because of the low differential range of the HAS even though the dynamic range covers 80 dB below ambient level. Frequency masking occurs when human ear cannot perceive frequencies at lower power level if these frequencies are present in the vicinity of tone- or noise-like frequencies at higher level.

Additionally, a weak pure tone is masked by wide-band noise if the tone occurs within a critical band. This property of in audibility of weaker sounds is used in different ways for embedding information. Embedding of data by inserting inaudible tones in cover audio signal has been presented recently[5].

In audio steganography, secret message is embedded into digitized audio signal which result slight altering of binary sequence of the corresponding audio file. The list of methods that are commonly used for audio steganography are listed and discussed below.

1.   LSB coding
2.   Parity coding
3.   Phase coding
4.   Spread spectrum
5.   Echo hiding.

### D. Video Steganography:

Video files are generally a collection of images and sounds, so most of the presented techniques on images and audio can be applied to video files too. When information is hidden inside video the program or person hiding the information will usually use the DCT (Discrete Cosine Transform) method. DCT works by slightly changing each of the images in the video, only so much that it is not noticeable by the human eye. To be more precise about how DCT works, DCT alters values of certain parts of the images, it usually rounds them up. For example, if part of an image has a value of 6.667 it will round it up to 7.

The great advantages of video are the large amount of data that can be hidden inside and the fact that it is a moving stream of images and sounds. Therefore, any small but otherwise noticeable distortions might go by unobserved by humans because of the continuous flow of information [5].

## 1.3. Image Steganography (Domain & its Techniques)

The most popular files for hiding data are the images. Image Steganography refers to the process of passing secret or confidential data in an image. In this process, an image is taken and secret message (payload) is set in that image and is passed to the sender. The sender can then extract the information from the image using the key provided by the sender.

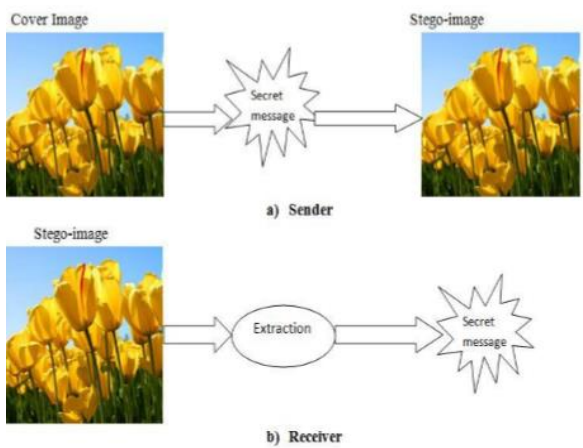

**Figure 3: Image Steganography Process**

Image Steganography Domain and Its Technique are:

A.  Spatial Domain
B.  Frequency Domain
C.  Masking and Filtering Domain

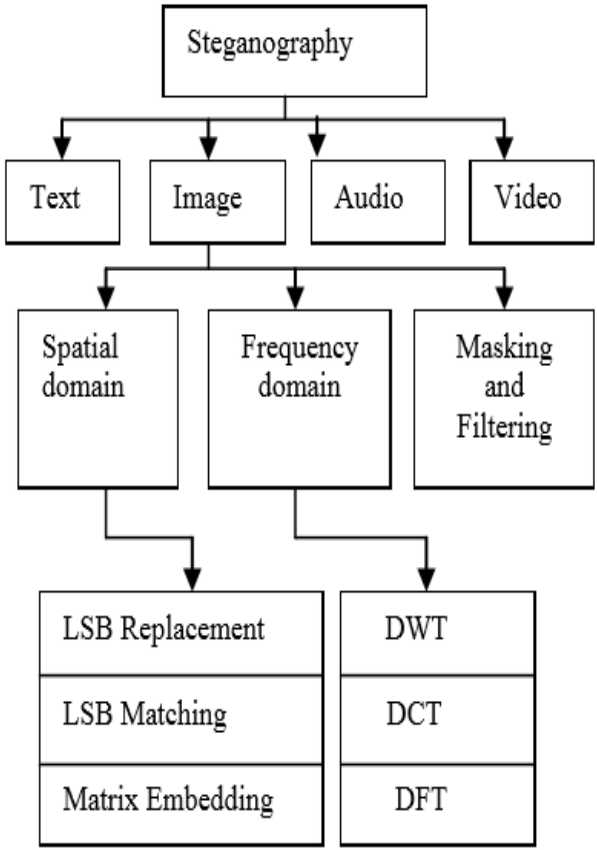

**Figure 4: Classification of Image Steganography**

### A.  Spatial Domain Steganography:

This method uses encoding in Least Significant Bits. LSB insertion method is an easy approach for embedding data into the actual image. There are numerous versions of this method; all these techniques reliably alter some of the bits in the values of image pixels for hiding data. LSB dependent steganography is one of the major techniques that hide confidential messages in the LSBs of some pixel values without any noticeable alterations. For our human eye, variations in the LSB are unnoticeable. Embedding of bits of data can be carried out either simply or randomly. LSB techniques as well as Matrix embedding are some spatial domain techniques [6].

### 1. LSB Replacement:

In this steganography, the cover pixel LSBs is substituted with a bit of the message that has to be embedded. Before embedding, the message is transformed into a sequence of bits, which are then inserted sequentially where the LSBs are located. This steganography is detectable even if there is low embedding rate.

Advantages of the LSB method are:
- Degradation of the original image is not easy
- Higher hiding capacity

Disadvantages of LSB method are:
- Robustness is low
- Simple attacks may destroy the embedded data

### 2. LSB Matching:

This type is much improved over LSB replacement method. In this process it is either randomly summed up or subtracted from the value of the cover pixel in case the bit of the confidential message is not equivalent to the LSB that come from the cover pixel. As compared to LSB Replacement method it is hard to detect LSB matching.

### 3. Matrix Embedding:

This technique encodes the original image as well as the message by an error correction code. It also alters the original image with respect to the result of coding. In this process the possible message bits are embedded randomly per an embedding change thus it helps in increasing embedding efficiency.

### B. Frequency Domain Steganography:

This is a more complex way of hiding information in an image. Various algorithms and transformations are used on the image to hide information in it. Transform domain embedding can be termed as a domain of embedding techniques for which a number of algorithms have been suggested. The process of embedding data in the frequency domain of a signal is much stronger than embedding principles that operate in the time domain. Most of the strong steganographic systems today operate within the transform domain Transform domain techniques have an advantage over spatial domain techniques as they hide information in areas of theimage that are less exposed to compression, cropping, and image processing. Some transform domain techniques do not seem dependent on the image format and they may outrun lossless and Lossy format conversions[7].

### 1. DWT (Discrete Wavelet Transformation):

Wavelets are described as the functions obtained over a fixed interval and have zero as an average value. This transformation is an extremely necessary way to be used for signal investigation as well as image processing, mainly for multi-resolution demonstration. It may crumble a signal into a number of constituents in frequency domain. 1-D DWT segments a cover image further into two major components known as approximate component and detailed component. A 2-D DWT is used to segment a cover image into mainly four sub components: one approximate component (LL) and the other three include detailed components represented as (LH, HL, HH).

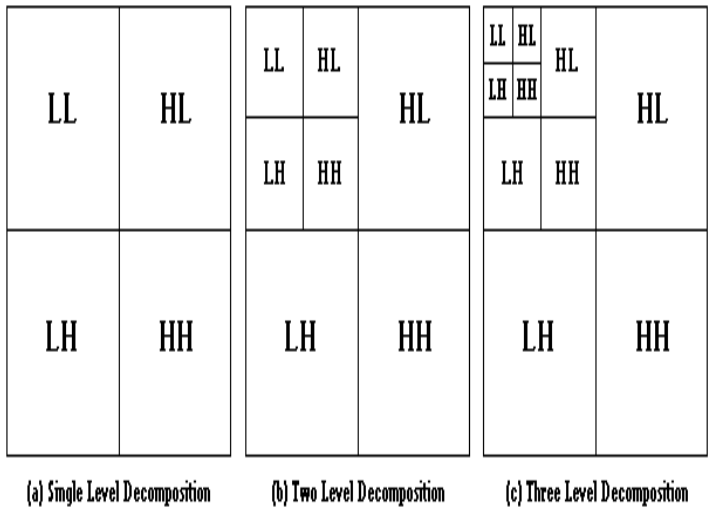

(a) Single Level Decomposition     (b) Two Level Decomposition     (c) Three Level Decomposition

**Figure 5: Discrete Wavelet Transformation Image Compression Level**

### 2. DCT (Discrete Cosine Transformation):

This transformation technique is useful for separating an image into different parts of differing significance (which is associated with the image's quality). It resembles the Fourier Transform Technique as it converts an image from its spatial domain into frequency domain. In this technique, for every color constituent, the JPEG format of image makes use of cosine transform to convert consecutive pixel blocks of size 8 x 8 into a count of 64 cosine coefficients each.

### 3. DFT (Discrete Fourier Transformation):

This technique is important as it separates an image into the sine and cosine values. It converts space and time dependent information into the frequency based information. It is useful for a number of applications including image filtering and reconstruction as well as image compression. It does not include all frequencies that result to form an image but constitutes of only the set of those samples, which are sufficient to describe the original image.

### C. Masking and Filtering:

These techniques hide information by marking an image, in the same way as to paper watermarks. These techniques embed the information in the more significant areas than just hiding it into the noise level. The hidden message is more integral to the cover image. Watermarking techniques can be applied without the fear of image destruction due to Lossy compression as they are more integrated into the image.

Advantages of Masking and filtering Techniques:

- This method is much more robust than LSB replacement with respect to compression since the information is hidden in the visible parts of the image.

Disadvantages of Masking and filtering Techniques:

- Techniques can be applied only to gray scale images and restricted to 24-b.
-

## 1.4. Steganography Tools

There are a number of tools available that automate the embedding of covert data within a cover medium. These tools range from open source, freeware and commercial tools. In this section we identify, discuss and compare open source or freeware tools especially. Some of the tools, which we investigated, have also steganalytic properties and functions, however, we discuss them aspects of data hiding only [15].

### A. **Silenteye:**

Silenteye is an open source, cross-platform and easy to use application. It can embed covert data in images (bmp, jpeg) and audio files (wav). With this tool, different steganographic and cryptographic algorithms can be used owing to its plug-in support.

### B. **Imagesteganography:**

Image Steganography is also a free software for hiding your information in image files. You can hide text message or files inside an image file. Just select the source file in which you want to hide the secret message, and then select the file to hide or write the text message to hide. Select the output image location and then click on start button to start encoding the file. Encoded image will have the secret message inside the image. You can use the decode option of the same tool to decode the hidden file or message from the image.

### C. **Hide 'N' Send:**

Hide'N' Send is also a small utility which offers steganography feature. It lets you hide any kind of file behind a JPG image file. It supports hashing and encryption too. Therefore, you can hide your date by encrypting. This adds an extra layer of security. Interface of the tool is simple and offers two tabs –one to hide data and other to extract data. You can select the options accordingly. Just run the tool, select the image file, then select the file which you want to hide, select the encryption type and then hide the data the image. Use the same tool again to extract the hidden information in the image.

### D. **Hallucinate:**

Hallucinate is a small software application developed specifically for helping you hide sensitive files inside images. The photos that carry the secret items can be opened just like any other pictures stored in your computer.It can be deployed on all Windows versions out there, provided that you have the Java working environment installed on the target computer.

### E.  Quickstego:

QuickStego is a lightweight encryption tool built specifically for helping you protect sensitive data from unauthorized viewing by hiding text messages in images.It sports a clean and straightforward layout that allows users to upload images into the working environment using the built-in browse function, so you cannot rely on the "drag and drop" support.QuickStego support file formats: BMP, JPG, and GIF, and lets you enter the text messages into a dedicated pane. Plus, you can import data from plain text files.The images with the hidden text can be saved to BMP file format.

### F.  Steganofile:

Steganofile is a System Utilities software allow users to hide a file in one or many host files, so it cannot be easily seen by other persons.It basically works by attaching the original file to the end of the host file. If more than one host file is used, the original file is splitted into the same number of host files. This have the advantage that each host file size might not increase substantially, so it cannot raise any suspicious. If a password is provided, the hidden file is encrypted with a basic scheme so more security will be added to the file.

## 2.  BACKGROUND STUDY & LITERATURE REVIEW:

The article [1] reviews about the different techniques used in digital steganography and also describing about the implementation of LSB method. It uses LSB insertion in order to encode data into a cover image. And an analysis of the performances is made using images ranging from 1:9 to 131 megapixels.

The paper [2] presents secure and high capacity based steganography scheme of hiding a large-size secret image into a small-size cover image.  Discrete Wavelet Transform (DWT) technique of transform domain has been used to scramble the secret message. And the performance has been investigated by comparing various qualities of the stego image and cover image.

In review paper [3], data hiding is done under JPEG Image by using Quantization Error Table (QET) resulting from processing the DCT image with quantization and DE quantization used for selecting the position for hiding secret bits in the image.

The review paper [4] describes about the stego tools and the comparison of different tools using the image as cover media for the performance measurement. It also introduces a robust and high payload steganographic algorithm.

The review paper [7] presents about the history steganography with its process comparing with cryptography. It also describes the communication system of these techniques. And also they improve the quality of image by using 12 bits instead of 8 bits.

The review paper [9] describes about the Image steganography and the image used to hide the data using some software and analyzing on the metadata of images after data hiding. The Review Paper [15] describes about the image Steganography and its detailed description about the techniques used for the same. They have analyzed various steganographic techniques and also have covered steganography overview its major types, classification, applications.

The Review paper [16] presents the Image Steganography and its Current techniques for hiding the data considering the modern application for communication with the comparison of Steganography, Cryptography and Encryption. The review paper [18] shows about image steganography, where edges in the cover image have been used to embed messages. It analysis that the more the amount of data to be embedded, larger the use of weaker edges for embedding.

In this paper [20] they provide a critical review of the steganalysis algorithms available to analyze the characteristics of an image, audio or video stego media and the corresponding cover media and understand the process of embedding the information and its detection. Also give a clear picture of the current trends in steganography so that we can develop and improvise appropriate steganalysis algorithms.

## 3.   COMPARISON OF STEGANOGRAPHY TOOLS:

In Table 1, the Steganography tools, which are frequently used for hiding text under cover, image are compared. The information given in this table was thoroughly researched. The table has been listed alphabetically by tool names. The table describes the comparison of tools on the basis of image features like Image size, Dimension, Concealed File and Image Formats with its algorithm applied.

All steganography tools are run with same text (Text message is:"Hello Friends! We love BMIIT. We believe in "Make it Happen through Innovations and Values. " ") and made a stego image with the help of same image. The difference between stego image and original image is shown in bellow diagram (Figure - 6).

Figure - 6 describe about the comparison of original image with stego image on the basis of pixel-by-pixel values. The difference can be recognized by the white pixels on the black area. The below analysis bring us to the conclusion that "Hallucinate" has the major pixel difference where as "Steganofile" has no visible difference.

Figure - 7 describe about the comparison of original image with stego image on the basis of histogram values. The difference can be recognized by change in the frequency of histogram values. The below analysis bring us to the conclusion that "Hide 'N' Send" has the major value difference where as "Steganofile" has no visible difference[15].

## Table 1: Steganography Tools Comparison

| Tool Name | Ref. | Conceale d Data Type | Stego Image properties | | | | | Additional Information |
|---|---|---|---|---|---|---|---|---|
| | | | Image Size (Increas ed by) | Dimensi on (Increas ed by) | Inpu t Imag e Files | Outp ut Imag e Files | | |
| Hallucinate | [1] | Any File Type | 3 Times | No change | Any | BMP , JPG | | - |
| Hide 'N' Send | [1][2] | Any File Type | 1 Times | Increase d | Any | JPG | | Algorithm used are LSB, FS, M-LSB, M-A and Password Protected |
| Image steganography | [1][3][15] | Any File Type | 12 Times | No change | Any | PNG | | - |
| Quickstego | [3][5][15] | Txt File | 15 Times | No change | Any | BMP | | - |
| Silenteye | [10][12] | Any File Type | 5 times | No change | Any | BMP , JPEG | | Algorithm used is LSB, Encryption, Password Protected, Adjustable Quality, Compression, Plugin support |
| Steganofile | [1][3][12] | Any | No Change | No change | Any | Any | | Password Protected |

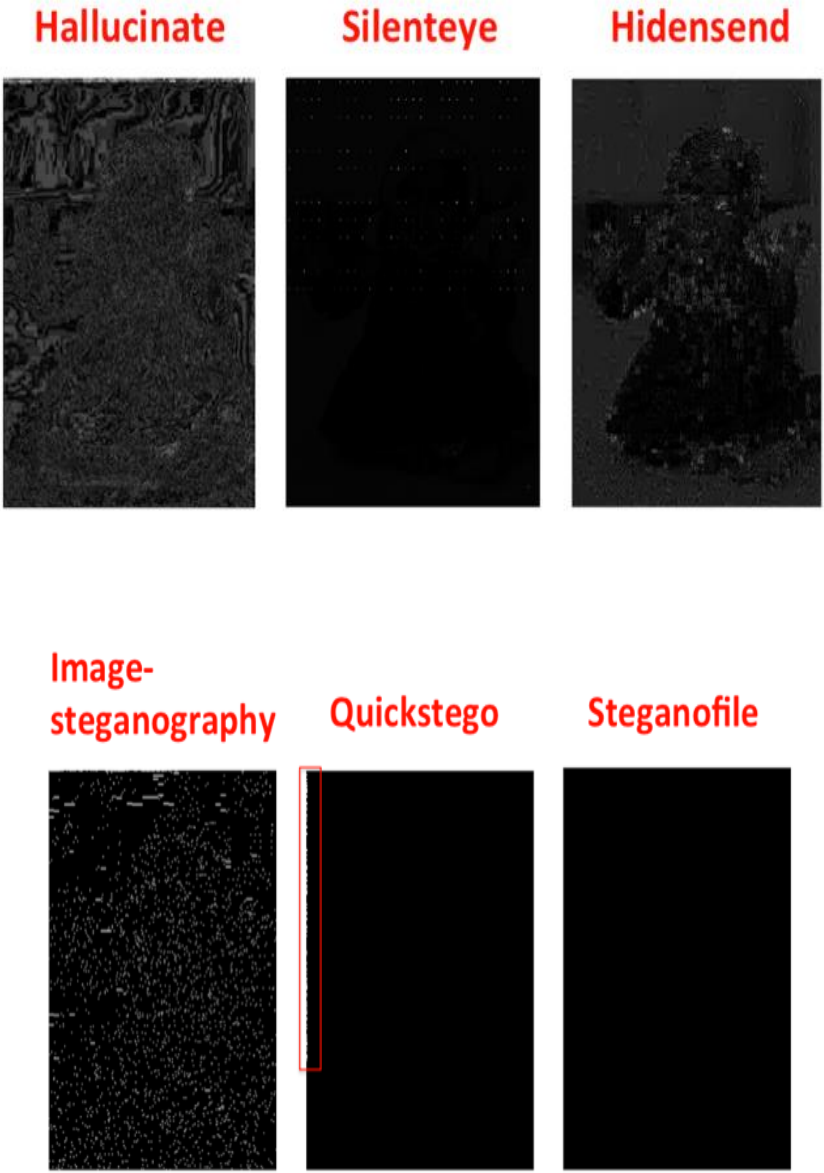

Figure 6: Pixel Value Differentiation of Original Image & Stego Image

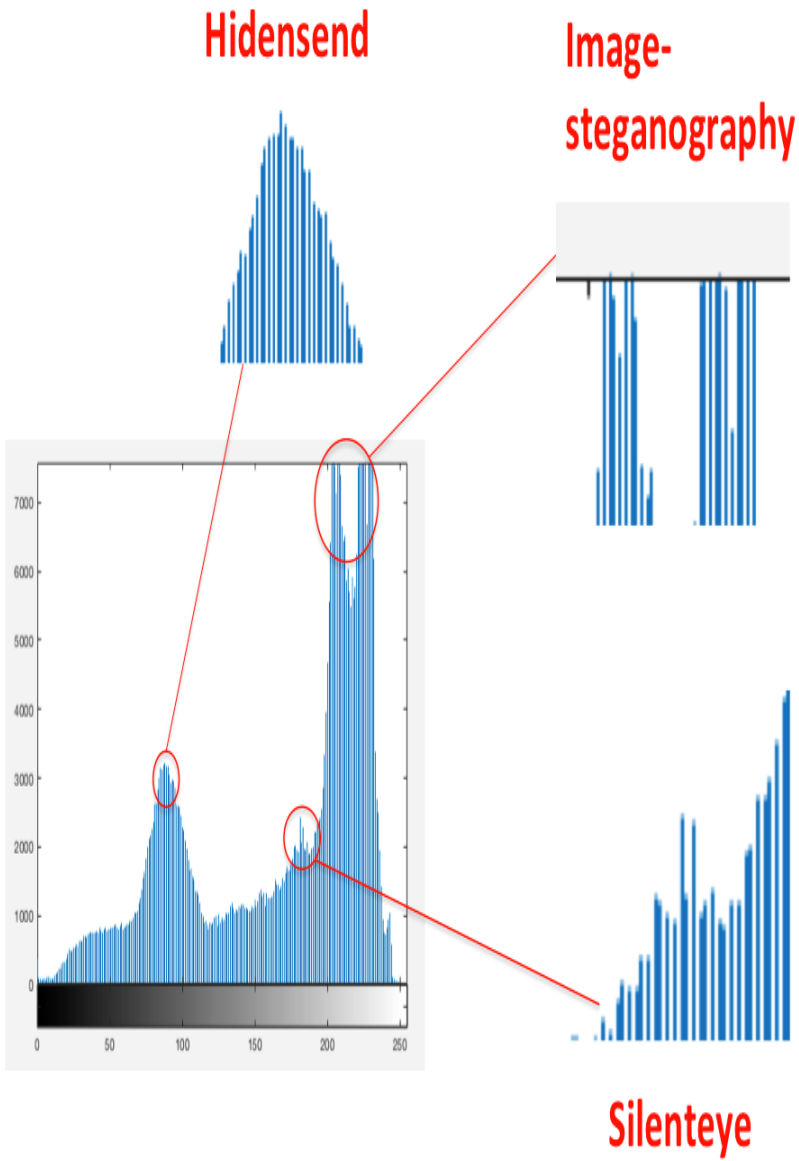

Figure 7: Histogram Differentiation of Original Image & Stego Image

## 4.  CONCLUSION

This research presents a background of Steganography and a comparative study of some Steganographic software. The attainment of this study is to identify the reliable and best tool available in the market for Steganography. Some of the tools available in the market were selected based on the frequent use; these tools were tested using the same input on all of them. Specific text was embedded within all host images for each of the six Steganography tools selected. The results of the experiment reveal that all the six tools were relatively performing at the same level, though some software performs better than others. Out of all the above Stego tools used for comparison, "Steganofile" (tool) was considered to be the most efficient one. This conclusion of efficiency was based on the image features like size, dimensions, and pixel value and histogram differentiation.

Steganofile has the advantages over all the other tools that it supports all the image formats and does not change the image features as well as does not reflect the visible changes.

## 5.   AUTHORS' PROFILE

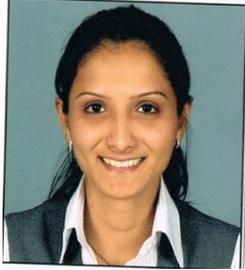

**Ms. Preeti Bhatt** is working as an assistant Professor at BMIIT. She has more than six years of experience in academics. Her interest area includes artificial intelligence. She has attended many seminar and workshops and also publishes research papers in national and international journal.   She   can   be   contacted   at preeti.bhatt@utu.ac.in

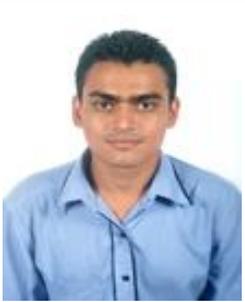

**Mr. Rakesh Savant** is working as an assistant Professor at BMIIT. He has more than three years of experience in academics. His interest area includes artificial intelligence, digital image processing and security. He has attended many seminar and workshops He can be contacted at rakesh.savant@utu.ac.in